\documentclass[fleqn, usenatbib]{mnras}

\usepackage{newtxtext,newtxmath}

\usepackage[T1]{fontenc}
\usepackage{ae,aecompl}

\usepackage{graphicx}	
\usepackage{amsmath}	

\usepackage{amssymb}	
\usepackage{bm}
\usepackage{soul}
\usepackage{footnote}

\def\Mdotin{\Mdot_{\mathrm{in}}}

\def\Mdot{\dot{M}}
\def\Edot{\dot{E}}
\def\Pdot{\dot{P}}

\def\Msun{M_{\odot}}

\def\Lcool{L_{\mathrm{cool}}}
\def\Lacc{L_\mathrm{acc}}

\def\rin{r_{\mathrm{in}}}
\def\rlc{r_{\mathrm{LC}}}
\def\rout{r_{\mathrm{out}}}
\def\rco{r_{\mathrm{co}}}

\def\Lx{L_{\mathrm{X}}}

\def\Md{M_{\mathrm{d}}}

\def\\rin{r_\mathrm{m}}
\def\v_esc{v_\mathrm{esc}}

\def\rA{r_{\mathrm{A}}}

\def\Tp{T_{\mathrm{p}}}

\def\Teff{T_{\mathrm{eff}}}

\def\Firr{F_{\mathrm{irr}}}

\def\P0min{P_{0,{\mathrm{min}}}}

\def\tc{\tau_{\mathrm{c}}}

\def\gpers{g~s$^{-1}$}
\def\ergpers{erg~s$^{-1}$}

\def\spers{s~s$^{-1}$}
\def\rinmax{r_{\mathrm{in,max}}}
\def\Rinmax{R_{\mathrm{in,max}}}
\def\reta{r_{\eta}}

\def\B0-P{$B_0$~--~$P$}
\def\PPdot{$P$~--~$\Pdot$}

\def\Torqdip{\Gamma_{\rm dip}}
\def\Torqacc{\Gamma_{\rm acc}}
\def\TorqD{\Gamma_{\rm D}}

\title[RRAT phases of INS systems]{Rotating Radio Transient Phases of Isolated Neutron Star Systems and Descendants of Central Compact Objects}

\author[Demirok et al.]{
Ş. Demirok$^{1}$,
Ü. Ertan$^{1}$\thanks{E-mail: unal@sabanciuniv.edu},
and A.A. Gençali$^{1}$
\\
$^{1}$Faculty of Engineering and Natural Sciences, Sabancı University, Tuzla, İstanbul, TR
}

\date{Accepted XXX. Received YYY; in original form ZZZ}

\pubyear{2025}

\begin{document}

\label{firstpage}
\pagerange{\pageref{firstpage}--\pageref{lastpage}}
\maketitle
\begin{abstract}
Rotating radio transients (RRATs) form an isolated neutron star (INS) population characterized by their sporadic and short radio bursts. 
In the fallback disc model, it was suggested that most INS populations pass through an RRAT phase during their evolution, and the mechanism producing these bursts could be the enhancement of the flux of open magnetic field lines through the poles of dead pulsars due to penetration of the inner disc into the light cylinder.  
To test this idea, we determine the RRAT phases that could be produced by this mechanism along the evolutionary curves of the INS families on the period-period derivative (\PPdot) diagram. 
These RRAT phases form an RRAT region in the $P$--$\dot{P}$~diagram, which is in agreement with the known RRAT distribution. 
Using our results, we also investigate the descendants of central compact objects (CCOs).  
We estimate that the CCOs similar to the three CCOs with measured $P$ and $\Pdot$ 
are not likely to become radio pulsars (RPs). 
The situation is rather different for Calvera, a recently identified CCO. 
In the model, Calvera-like sources become RPs with lifetimes $ > 10^{8}$~yr. Indeed, there is a cluster of RPs in the region of \PPdot~diagram where descendants of Calvera-like CCOs evolve in our model. We also estimate that they have a very low birth rate, and therefore their immediate descendants may not currently exist in our Galaxy.
\end{abstract}




\section{Introduction} \label{intro}

Rotating radio transients (RRATs) are isolated neutron star (INS) systems that show repetitive  radio bursts with durations in the $0.5-100$ ms range, much shorter than their spin periods. The time intervals between these radio bursts vary from minutes to hours.  
At present, there are $336$ known RRATs \citep{Agarwal2026} \footnote{For the recent data, see https://rratalog.github.io/rratalog.}.
For $230$ RRATs, only spin periods, $ P$, while for $54$ sources, both $P$ and period derivatives, $\Pdot$ , were measured through phase-coherent timing analysis over long-term observations \citep{Keane2011a,Keane2011b,Agarwal2026}. 
Their $ P$ and  $\Pdot$ values are in the ranges of $0.1 < P < 18$ s and $1.9 \times 10^{-17} \lesssim \Pdot \lesssim 5.6 \times 10^{-13} $ \spers~respectively \citep{Agarwal2026}.
The observed radio flux densities of RRATs vary between $12~-~3600$ mJy \citep{Agarwal2026}.
Among these RRATs,  only one source was detected in X-rays with an X-ray luminosity $\Lx \simeq 4 \times 10^{33} $ \ergpers ~\citep[PSR J1819--1458][]{Rea2009}, and there are $\Lx$ upper limits for some sources \citep{Rea2008,Kaplan2009, Keane2013}. 
The overall lack of X-ray detections might be due to uncertainties in their positions \citep{Kaplan2009}.

The physical mechanism that produces these bursts is not well understood \citep{Mc2006}. 
Different mechanisms were proposed to explain RRAT emission. It was suggested that RRATs could be ordinary radio pulsars with extreme nulling \citep{Wang2007nulling,Zhou2023}. 
It was shown that transitions between different quasi-stable magnetospheric states could account for nulling and mode changing in pulsars \citep[][see also \citealt{Jones2013}]{Timokhin2010,Li2012}. 
It was also proposed that RRATs could be distant radio pulsars with highly modulated pulse-amplitude distributions, which would have been detected as regular radio pulsars if they were closer \citep[][see also \citealt{Zhou2023}]{Weltevrede2006}.
The presence of circumstellar material around RRATs was also studied in different work. It was suggested that the material entering the light cylinder could inject charges into the magnetosphere and thereby alter the pair production, triggering the RRAT bursts \citep{Li2006, Cordes2008}. The mechanism adopted in the present work also relies on the presence of a disc inside the light cylinder, but the bursts are produced by the enhancement of the voltage across the poles of the NS rather than by charge injection into the magnetosphere \citep{Gencali2024}.

It was shown by a series of earlier work that $\Lx$ and rotational properties of INS populations could be achieved by NSs evolving with fallback discs. 
In the fallback disc model \citep{Chatterjee2000,Alpar2001,ErtanEtAl2009}, distinguishing properties of these INS families, namely, anomalous X-ray pulsars (AXPs), soft gamma repeaters (SGRs), dim isolated NSs (XDINs), high-magnetic-field RPs (HBRPs), central compact objects (CCOs), RRATs, and long-period transients (LPTs) emerge naturally as a result of their differences in the initial conditions (magnetic dipole moment, the initial period, and the disc mass) of their long-term evolution  \citep[see e.g.][ and references therein]{Gencali2024}. 

Fallback discs are expected to form around NSs following supernova explosions \citep{Colgate1971,Chevalier1989,Mineshige1993,Perna2014}.
Fallback disc model \citep{Chatterjee2000, Alpar2001} was proposed to account for
the $P$ clustering and the $\Lx$ values of AXPs. 
\citet{Alpar2001} proposed that the emergence of not only AXPs but also other INS populations can be explained if the properties of fallback disc are included in the initial conditions together with $P_0$ and $B_0$. 
Emission properties of fallback discs were studied extensively \citep{Perna2000, ErtanCaliskan2006}. 
In particular, the broadband spectrum of the AXP 4U~0142+61 from optical to mid-infrared \citep{Hulleman2000, Hulleman2004, Morii2005, Wang2006} has been reproduced by emission from an X-ray irradiated, viscously active gaseous fallback disc \citep{Ertan2007}. 
Disc emissions were also  discussed for XDINs, based on their optical excess \citep{Ertanetal2017} and extended infrared emission \citep{Posselt2018}.
Properties of hard and soft X-ray spectra of AXP/SGRs  were explained by the emission from the accretion column and the poles of the NS \citep{Trumper2010, Trumper2013, Kylafis2014, Zezas2015}. 

The fallback-disc model was further developed by \citet{ErtanEtAl2009} to study the long-term evolutions of AXP/SGRs and other NS populations 
\citep{Ertan2014,Benli2016,Benli2018HBRPs, Benli2018CCOs,Gencali2021RRATs, Gencali2022_GLEAM-X, Gencali2023, Gencali2024}.
Recently, details of the evolutionary links between INS populations were investigated by \citet{Gencali2024}. 
With this model, they found that RRATs, which are estimated to have highest birth rate \citep{KeaneKramer2008,Keane2011a}, have evolutionary links with most of the other INS populations. 
This provides a concrete support to the ideas proposing that evolutionary links could account for the birth rate problem, the discrepancy between the estimated cumulative INS birth rate and the core-collapse supernova rate (CCSN) in the Galaxy \citep{KeaneKramer2008,Keane2011a}.

A rotation-powered pulsar is expected to switch off its ordinary pulsed radio emission when the voltage across its polar cap falls below the critical level required to sustain the electron-positron pair production \citep{Chen1993}. 
Different magnetic field geometries define a death valley on the \PPdot~ diagram \citep{Ruderman1975,Chen1993}. Each pulsar has its own death point within the valley. 
When the source reaches this point, radio pulsations are expected to cease, and the source becomes a “dead pulsar".  In an earlier work, in a different context related to millisecond pulsars in LMXBs, \cite{Parfrey2016} proposed that the penetration of the accretion disc into the light cylinder increases the flux of the open field lines through the poles of NS, and thereby enhances the  voltage across the poles. 
This enhancement depends sensitively on the location of the inner disc radius, $\rin$, relative to the light cylinder radius, $\rlc$.
\cite{Gencali2024} proposed that this mechanism could be responsible for the emission of RRAT bursts from systems that would be dead pulsars without this voltage enhancement.  
This means that, for a given INS family, there could be a certain evolutionary phase during which these sources could show RRAT behaviour. We will call these epochs "RRAT phases".

In this work, we test this idea by comparing the results of the model calculations with the rotational properties of RRATs in the \PPdot~diagram.  
Based on the results of this analysis, we will also investigate possible descendants of CCOs in the same model, and compare our results to the predictions of other models assuming that CCOs evolve in vacuum purely by dipole torques \citep[see e.g.][]{Gotthelf2013DRP,Luo2015}. 
CCOs are INSs observed at the centers of supernova remnants (SNRs). They emit only thermal emission with $\Lx \sim 10^{33}$~\ergpers. Among 10 confirmed CCOs, three sources have measured $P \sim 0.1-0.4$~s and $\Pdot \sim (0.73 - 2.2) \times 10^{-17}$\spers. 
Their observed $\Lx$ values are one to two orders of magnitude higher than their spin-down powers. Their SNR ages are estimated to be in the $0.33 - 27$ kyr range \citep[see][for a review]{DeLuca2017}.
In Section~\ref{the model}, we briefly describe our model. 
We discuss our results in Section~\ref{results}, and summarize our conclusions  in Section~\ref{conclusions}.

\section{The MODEL} \label{the model}

\subsection{Evolution of NSs with  Fallback Discs}

Here, we briefly describe the long-term evolution model previously applied to different INS populations  \citep[see e.g.][]{Ertan2014,Benli2016,Gencali2022_GLEAM-X} 
and the analytical model used for the $\rin$ and torque calculations \citep{Ertan2017,Ertan2018,Ertan2021}. 
We solve the disc diffusion equation for a geometrically thin and optically thick accretion disc using the $\alpha$ prescription for the kinematic viscosity, 
$\nu = \alpha h c_s,$ where $\alpha$ is the viscosity parameter, $c_s$ is the sound speed, and $h$ is the pressure scale height of the disc \citep{Shakura1973}.
The disc is heated by viscous dissipation and X-ray irradiation.  The X-ray irradiation flux can be written as $\Firr \simeq 1.2\ C\, \Lx / (\upi r^2)$ where $r$ is the radial distance of the disc, $C$ is the irradiation efficiency parameter \citep{Fukue1992}.
The total X-ray luminosity $\Lx =\Lcool + \Lacc$ where 
$\Lcool$ is the cooling luminosity \citep{Page2006,Page2009},
$\Lacc = G M \Mdot_* / r_*$ is the accretion luminosity, $G$ is the  gravitational constant, $M$ and $r_*$ are the mass and radius of the NS, and $\Mdot_*$ is the mass accretion rate on to the NS. 
We note the difference between the mass inflow rate of the disc  $\Mdotin$ and $\Mdot_*$.
$\Mdotin$ is calculated from the solution of the diffusion equation for each time step \citep{FrankKingRaine2002}. We take $\Mdot_* = \Mdotin$ when accretion is allowed and  $\Mdot_* = 0$ otherwise (see below).
The effective temperature of the disc $\Teff \simeq \left[ (D + \Firr)/\sigma \right] ^{1/4} $, where $D$ is the  viscous dissipation rate, $\sigma$ is the Stefan–Boltzmann constant.
When the local $\Teff$ drops below a critical temperature, $\Tp$, the disc becomes viscously inactive \citep{ErtanEtAl2009}. 
The dynamical outer radius, $\rout$,  of the active disc is the radius at which currently  $\Teff=\Tp$. During the long-term evolution, $\rout$ gradually decreases with decreasing $\Firr$, and eventually the entire disc becomes viscously inactive. 
The disc parameters ($\alpha$, $\Tp$, and $C$) are expected to be similar for the fallback discs of different INS families. 
For the earlier applications of the model, we obtained reasonable results  with $\alpha = 0.045$, $\Tp = 50$--$150$~K, and $C = (1$--$7)\times10^{-4}$  \citep[see e.g.][]{ErtanCaliskan2006,Caliskan2013,Benli2017, Benli2018CCOs,Gencali2021RRATs}.

In the fallback disc model, the initial conditions of the sources are  the initial period, $P_0$, the dipole field strength on the poles of the NS, $B_0$, and the initial disc mass, $\Md$.
Among these, except very young sources, the long-term evolution of the sources is found to be most sensitive to $B_0$ \citep[see ][for a detailed study investigating the effect of model parameters and initial conditions on the long-term evolution]{Benli2016}.

We determine $\rin$, torques, and the critical conditions for the transitions between the rotational phases of the NS using our analytical model \citep{Ertan2017,Ertan2018, Ertan2021}. 
This model is based on the basic principles of the model developed by \citet{Lovelace1995} and \citet{Ustyugova2006} \citep[see e.g.][for a detailed explanation]{Ertan2017}.
In the model developed by \citet{Ertan2021}, there are three rotational phases, namely, strong propeller (SP), weak propeller (WP), and spin up (SU) phases. The location of $\rin$  relative to the co-rotation radius, $\rco = (GM/\Omega_*^2)^{1/3}$, where $\Omega_*$ is the angular velocity of the NS, determines the current rotational phase of the source. There is not a single $\rin$ equation valid for all rotational phases. Below, we briefly describe how $\rin$, rotational phases, and the torque acting on the NS vary for an illustrative model source with gradually increasing $\Mdotin$. 

For low $\Mdotin$ rates, the source is in the SP phase during which all the inflowing disc matter is thrown out of the system from the narrow inner-disc boundary with radial width $\Delta r < r$. Outside the boundary, the field lines are open and disconnected from the disc \citep[see e.g.][]{Lovelace1999,Uzdensky2004,Ustyugova2006}.
In this phase, the maximum radius at which the SP mechanism is sustained can be written as 
\begin{equation}
\Rinmax^{25/8} \left| 1 - \Rinmax^{-3/2} \right|
\simeq 1.26 \, \alpha_{-1}^{2/5} \, M_{1.4}^{-7/6} \,
\Mdot_{\rm in,16}^{-7/20} \, \umu_{30} \, P^{-13/12}
\end{equation}
\citep{Ertan2021}. 
Here, $\Rinmax =\rinmax/\rco$, 
$\alpha_{-1} = (\alpha/0.1)$, 
$M_{1.4} = (M/1.4\,M_\odot)$, 
$\umu_{30} = \umu/(10^{30}\,{\rm G\,cm^3})$ is the magnetic dipole moment. 
In all our calculations, we keep $\mu$ constant. When the sources are in the SP phase, the field decay for conventional dipole fields in our model is negligible for $\sim 10^6$~yr \citep{Vigano2013unifying}. Field decay could be significant during the early evolutionary phases for the accreting sources with relatively high $\Mdot_*$. In the model, for most sources starting evolution in the WP phase, we estimate from the simulations that $\Mdot_*$ decreases below $\sim 10^{13}$~\gpers~at ages of a few $10^4$~yr. Afterwards, the field decay is not significant due to sharply decreasing $\Mdot_*$ \citep{Konar1997,Konar1999,Konar2017}. 
Initial dipole fields of these sources could decay to a few times their initial  $B_0$ mostly during the early phase (a few $10^4$~yr) of their evolution.  
For these sources with such field decays, similar source properties could be achieved at ages smaller than the age estimated with constant $B_0$.
We define $\reta = \eta \rinmax$ and the actual inner disc radius in the SP phase is estimated to be $\rin = \reta$ with an $\eta \lesssim 1$.  
Speeds of the closed magnetic field lines co-rotating with the star exceed the escape speed beyond $r_1 = 1.26~\rco$.
This means that a steady SP phase can be sustained with $\rin > r_1$. 
In this phase, there is no mass accretion on to the NS ($\Mdot_* = 0$) and  $\Lx = \Lcool$.
At relatively higher $\Mdotin$ levels, if $\rin$ is instantaneously between $\rco$ and $r_1$, the disc matter  expelled from the inner disc boundary returns to  disc causing a pile-up at larger radii. This pushes the inner disc inwards until $\rin =\rco$, which starts the accretion on to the NS taking the system into the WP phase.
Due to weak $\Mdotin$ dependence of $\rin$ in this model, the WP phase can persist for a wide range of $\Mdotin$ with $\rin =\rco$. 
In this phase, $\Mdot_* = \Mdotin$ and $\Lx = \Lacc + \Lcool$.

The disc can penetrate inside $\rco$ when  $\Mdotin$ exceeds a critical level at which the viscous and magnetic stresses are balanced. 
This corresponds approximately to the rate that satisfies $\rco \simeq r_{\xi} = \xi \rA$ where $\rA$ is conventional Alfvén radius, and $\xi$ is a parameter usually estimated to be in the $0.5 \lesssim \xi \lesssim 1$ range \citep{Ghosh1979,Wang1995, FrankKingRaine2002}. This also corresponds to the WP/SU transition (torque reversal) of the system \citep[see][for details]{Gencali2022_4U1626}. Since most INS populations do not enter the SU phase in this model, we will not discuss the details of this phase here. 

The total torque, $\Gamma$, acting on the system can be written as
\begin{equation}
\Gamma = \Mdot_* \sqrt{G M \rin}
- \frac{\mu^2}{\rin^{3}}\left(\frac{\Delta r }{ \rin}\right)
- \frac{2}{3} \frac{\mu^2 \Omega_*^3}{c^3}
\end{equation}
where $c$ is the speed of light. The first term on the right hand side is the accretion torque,  $\Torqacc$, resulting from the mass accretion on to the NS. The second term is the disc torque, $\TorqD$, produced by the disc-field interaction. The last term is the magnetic dipole torque, $\Torqdip$, which is mostly negligible compared to $\TorqD$ and $\Torqacc$. For the calculation of $\Torqdip$, we take the magnetic obliquity angle $90\degr$, which maximises the dipole torque.
The spin evolution is obtained from $\Gamma = I\dot{\Omega}_* = -2\upi I \Pdot P^{-2}$ where $I$ is the moment of inertia of the NS. 
In the WP phase, all the torques are active
while $\TorqD$ mostly dominates the other mechanisms. 
In the SP phase, $\Mdot_* = 0$ and $\Gamma = \TorqD + \Torqdip$.
After the entire disc becomes inactive, the NS spins down with relatively weak $\Torqdip$ alone. 
In this model, the accretion on to the NS ceases the ordinary radio pulsations,    
while the radio pulses are  emitted during the SP phase, provided that the source has a sufficient power.  
This is because the open field lines have no connection with the disc outside the inner boundary \citep[see e.g.][]{Ustyugova2006}.       



\subsection{Disc Induced Voltage Enhancement in NSs} \label{parfrey's effect}

\begin{figure*}
    \centering
    \includegraphics[width=\linewidth]{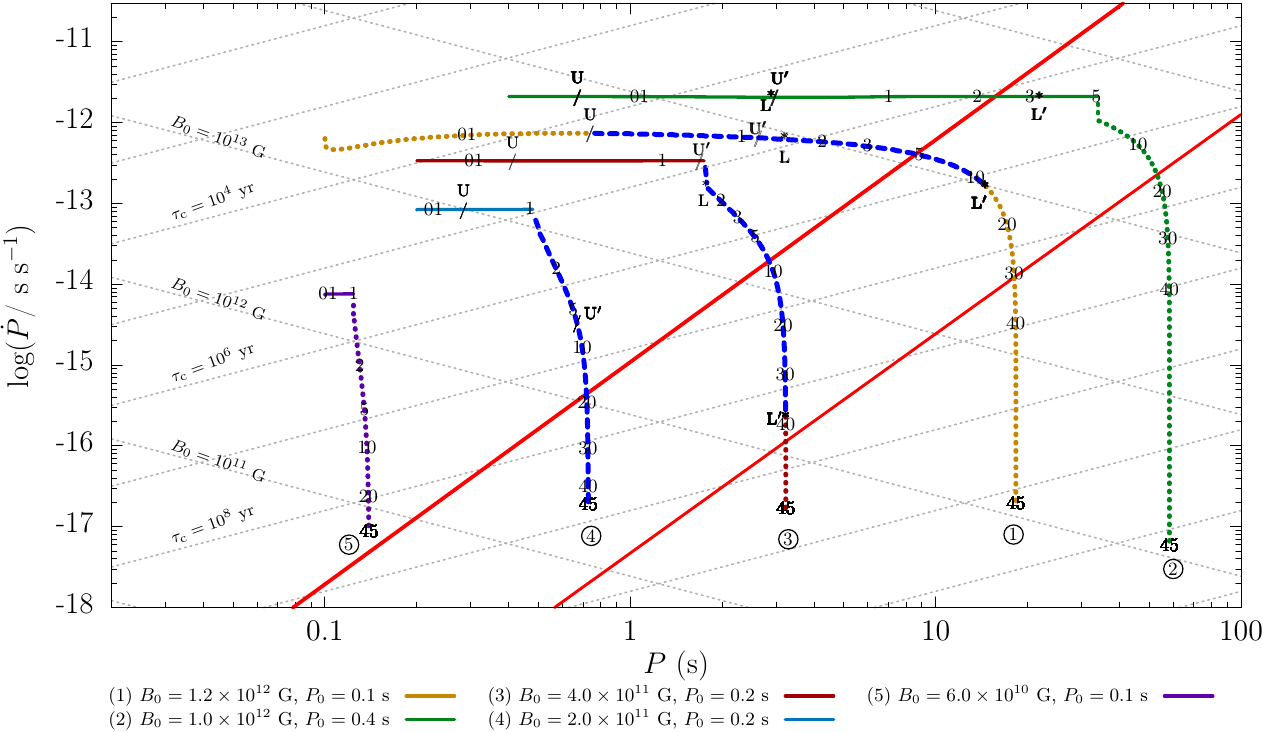}
    \caption{Illustrative model curves that show different  evolutionary paths. 
    For all these model curves: $\alpha = 0.045$, $\Delta r/\rin = 0.25$, $\eta = 0.7$, $\xi = 0.7 $, $\Tp = 50$~K, and $C =1 \times 10^{-4}$.
    The initial conditions $B_0$ and $P_0$ are given below the figure, and $\Md \simeq 10^{-5} \Msun$ for all model curves.
    The solid and dotted segments of the model curves represent the WP and SP phases respectively.
    Diagonal, solid lines show the borders of the pulsar death valley \citep{Chen1993}. Diagonal, dotted lines show constant $\tc$ and $B_0$ lines calculated from the dipole-torque formula.
    The numbers in circles are given to identify each curve for the discussion in the text. The numbers on the model curves show the ages of the sources in units of $10^5$ yr. 
    The points $U$ (/) and $L$ ($\ast$) on the curves indicate the periods at which the sources cross the upper and lower boundaries of the death valley with the $B_0$ value in our model. The disc-induced voltage enhancement shifts the points $U$ and $L$ to $U^{\prime}$ and $L^{\prime}$ respectively.
    Along a model curve, in the SP phase, RPs could be observed until $L$, and RRAT bursts could be detected between $U$ and $L^{\prime}$. After $L^{\prime}$, no radio emission is expected, while the disc could remain active (until the end of the model curve). See the text for details.
}
    \label{fig:PPdot_model}
\end{figure*}

When the inner disc penetrates into the light cylinder, the closed field lines open up down to $\rin$. This  enhances the flux of the open magnetic field lines through the poles of the NS which could be written as 
\begin{align}
\Phi = \zeta ~~ \frac{\rlc}{\rin} ~~ \Phi_0 
\end{align}
\citep{Parfrey2016, Parfrey2017}, where $\Phi_0 = \upi r_*^2 (\Omega_* r_* / c) B_0$ is the flux for a vacuum source (without disc enhancement), $\zeta$ is a parameter estimated to be close to unity, and $\rlc = c/\Omega_*$ is the light cylinder radius. In our calculations, we take $\zeta = 1$.
With this enhanced $\Phi$, the maximum potential across the polar cap becomes
\begin{align}\label{voltage}
\Delta V_{\text{max}} 
\approx \frac{B_0 r_*^3 \Omega_*^2}{2c^2} 
= \frac{\Omega \Phi}{2\upi c}
\end{align}
\citep{Ruderman1975}. 
For a NS rotating in vacuum, the critical condition for the pair production that  can yield the radio emission is given by
\begin{align} \label{condition}
\left( \frac{e \Delta V}{m c^2} \right)^3 \frac{\hbar}{2 m c r_{\rm c}} \frac{h_{\rm p}}{r_{\rm c}} \frac{B_{\rm s}}{B_{\rm g}} \approx \frac{1}{15},
\end{align} 
\citep{Chen1993}, where $e$ and $m$ are the  charge and mass of an electron,  $\hbar$ is the reduced Planck constant, $h_{\rm p}$ is the thickness of the polar cap gap, $r_{\rm c}$ is the curvature radius of the magnetic field line, $B_{\rm g} = m^2 c^3/(e \hbar) = 4.4 \times 10^{13}$ G is the quantum critical magnetic field strength, 
and $B_{\rm s}$ is the local magnetic field strength at the pole of the NS, which could be significantly different from the actual $B_0$ value. 
For different magnetic field geometries, \citet{Chen1993} estimated the lower and upper boundary of the pulsar death valley for the NSs evolving in vacuum, using equations (\ref{voltage}) and (\ref{condition}) (diagonal solid lines in Fig. \ref{fig:PPdot_model}). 
RPs are expected to die at a point inside this death valley.
This means that each RP has its own death point inside the valley. 
These death points seem to be randomly distributed within the valley considering that there is no clustering inside the valley \citep{Zhang2000, Beskin2022}.

It was proposed earlier that the disc-induced voltage enhancement of dead RPs could be the mechanism yielding the RRAT bursts \citep{Gencali2024}.
For a given $P$, this voltage enhancement effect is equivalent to decreasing the critical $B_0$ 
for the radio emission (in the form of RRAT bursts) on the \PPdot~diagram. 
This means that the lifetime of a RP could be significantly extended with repetitive short radio bursts, rather than regular radio pulsations 
as a result of the disc induced voltage enhancement.  
In this model, most RRATs are estimated to evolve in the SP phase \citep{Gencali2021RRATs}.
In this phase, inflowing disc matter is expelled from the narrow inner disc boundary. The closed field lines interacting with matter inflate, open up, and reconnect on the dynamical timescale,  
while the matter could escape the system along the open field lines \citep[see e.g.][]{Lovelace1999,Ustyugova2006}. 
We estimate that this continual interaction between the field lines and the inner disc 
could produce a transient radio emission with very short durations. 
The details of this transient radio emission mechanism is beyond the scope of the present work.

In Eq. \ref{condition}, $h_{\rm p} \simeq r_*(r_* \Omega_* /c)^{1/2}$, $r_{\rm c} = (r_* c / \Omega_* )^{1/2}$ and $B_{\rm s} =B_0$ for the upper boundary. Considering $\Delta V_{\text{max}}$ is enhanced by $\rlc/\rin$, the critical $B_0$ required for the  radio emission decreases by a factor of $(\rlc/\rin)^{3/4}$ for the current $P$ of the NS. For the lower boundary, $h_{\rm p} \simeq (B_0/B_{\rm s})^{1/2}r_*(r_* \Omega/c)^{1/2}$, $r_{\rm c} = r_*$ and $B_{\rm s} =2 \times 10^{13}$~G \citep{Chen1993}, which  decreases  the critical $B_0$ by a factor of $(\rlc/\rin)^{6/7}$.
At each time step, in addition to our model calculations, we also calculate the critical $B_0$ values for RPs (with dipole torque only) corresponding to each border of the death valley for the current $P$ values. Along each model curve, we determine the two critical points ($B_0$, $P$) at which the curve crosses the borders \citep[calculated by][]{Chen1993}. For the same source, we also determine the critical points estimated with the disc enhancement. From these calculations, we estimate the evolutionary phases during which NSs with fallback discs could show RRAT bursts in this picture (see below).

We will test this idea with the observed rotational properties of RRATs as follows. A given RP evolving in vacuum  could die between two extreme points on its evolutionary curve: on the upper and lower borders of the death valley in the \PPdot~diagram. 
Let us denote these points by $U$ and $L$ respectively. 
Let us consider an imaginary non-accreting source with a death point exactly on point $U$. The voltage enhancement by the disc extends its radio-active lifetime as an RRAT to a later point on its evolutionary curve, which we denote by $U'$.
This source shows regular radio pulses until it crosses the point $U$ (follow e.g model curve 3 in Fig.~\ref{fig:PPdot_model} for the explanation below). 
Subsequently, it becomes an RRAT between the points $U$ and $U'$. The source shows no radio activity after the point $U'$. Along the same evolutionary curve, consider another source on the other extreme, with a death point on the lower border, that is, on point $L$. Similarly, this source evolves as an RP until it crosses the point $L$, and as an RRAT between $L$ and $L'$. Along a given evolutionary curve, assuming a random distribution of RP death points between $U$ and $L$, we could observe RRATs between the points $U$ and $L'$. In the $P$--$\dot{P}$ diagram, the $U$ and $L$ points along the model curves are located much above the upper and lower borders, respectively due to presence of disc torques. For a given $P$ and $B_0$, disc torques yield $\dot{P}$ values much greater than those generated by dipole torques only. A model source at point $U$ would be on the upper border if it were evolving only with dipole torques (see the example below). To sum up, along a model curve, there are: (1) only RPs before $U$, (2) both RRATs and RPs between $U$ and $L$, (3) only RRATs between $L$ and $L'$, and (4) no radio activity after $L'$ even if the disc could still be active (until the end points of the model curves).

For a given source, there may not be a sharp transition from the RP to the RRAT emission. During this transition, the NS could show both radio pulsations and bursts. Indeed, PSR B0656$+$14 emits bright radio bursts together with much weaker radio pulses, and it would have been detected as an RRAT if it were more distant \citep{Weltevrede2006}. 
This led to the prediction that at least some RRATs may emit weaker pulses in addition to their occasional bright bursts \citep{Weltevrede2006}. This prediction was confirmed by the sensitive observations of RRATs by \citet{Zhou2023} with the Five-hundred-meter Aperture Spherical Radio Telescope (FAST). Out of 59 observed RRATs, weak regular radio pulses were observed from 25 sources \citep{Zhou2023}. We note that we do not take into account the factors that can prevent the detection of the radio pulses or bursts, such as the beaming effect, decreasing radio luminosity with decreasing $\Edot$, and the alignment of the rotational and magnetic axes, which should be considered in  statistical analyses. 
In the model, the mass-flow on to the NS hinders not only regular radio pulses of RPs but also radio bursts of RRATs.
In our model, since $\rin$ is a function of $P$ as well as $B_0$ and $\Mdotin$, we should determine the points $U'$ and $L'$ for each evolutionary curve separately. In other words, there is no "RRAT valley" in this model. 
Instead, 
we will obtain a RRAT region in the \PPdot~diagram formed by the RRAT phases of the model curves 
for different populations and compare it with the distribution of RRATs with known $P$ and $\Pdot$ values.

We will explain the calculation of the initial and final points of the RRAT phases using the illustrative model curves given in Fig.~\ref{fig:PPdot_model}. 
For the disc parameters $\alpha$, $\Delta r/\rin$, $\eta$, and $\Tp$ (see Fig.~\ref{fig:PPdot_model}), we have used the typical values that gave reasonable results in our earlier work \citep[see e.g.][]{Ertan2018, Ertan2021, Gencali2024}.
Constant $B_0$ lines seen on the \PPdot~diagram are calculated from the dipole torque formula for the sources rotating in vacuum. The $B_0$ values estimated in our model given below the figure are very different from those indicated by the constant $B_0$ lines, which are given for comparison.
Along a given model curve, $B_0$ remains constant. 
The numbers on the curves indicate the source ages in units of $10^{5}$ yr. 
Along the evolutionary paths, the solid lines correspond to the WP phases during which accretion on to the sources quenches both normal radio pulses and RRAT emission. 
The dotted curves show SP phases during which radio pulsations are allowed if the sources have sufficient power. 
During the SP phase, a RP could make a transition to the RRAT phase at a point between $U$ and $L$ depending on its death point inside the death valley. 
RRATs could be observed between  $U$ and $L'$.
The dashed segments of the model curves show the RRAT phases all of which lie in the SP phases.
A large fraction of sources start their evolution in the WP phase, and transition to the SP phase at a later time (curves 2--5), while for the remaining fraction of the sources the evolution starts in the SP phase (curve 1). 
Among these, some sources could start the RRAT phase immediately after the WP/SP transition (curves 3 and 4), while some other sources can never enter the RRAT phase (curve 2). 
For the evolutions illustrated by the curves 1 and 3, RRAT phase could terminate while the disc is still active. 
There are also initial conditions leading to RRAT phases terminating with the inactivation of the disc (curve 4). 
It is also possible that the disc could become inactive while the source is a RP and still above the upper boundary (curve 5), which is likely to evolve as a RP, and never enter the RRAT phase. 
Note that this source cannot reach the $U$ point while its disc is active.
For sources 1 and 3, the SP phase goes on after $L'$ points, while the sources do not have sufficient power for RP or RRAT emission.
For any curve, along the RRAT phase (dashed segments), there could be both RRATs and RPs between $U$ and $L$ points, while only RRAT emission is possible between $L$ and $L'$.

We can further clarify the meaning of $U$ and $L$ points using one of the illustrative curves in Fig. \ref{fig:PPdot_model}. 
For example, for the curve 2 with $B_0 = 1 \times 10^{12}$~G, the point U is at $P \simeq 0.7$~s.
Note that the constant $B_0 = 1 \times 10^{12}$~G line crosses the upper border at the same period. 
At point $U$ on this curve, $\Pdot \simeq 2 \times 10^{-12}$~ \spers~is a few orders of magnitude greater than  $\Pdot \simeq 4 \times 10^{-16}$~\spers~on the upper border for the same $P \simeq 0.7 $ s. This is due to relatively efficient disc torques.
This means that a source currently at point $U$ on curve 2 would be on the upper border of the valley at the point corresponding to $P \simeq 0.7$ s if there were no disc torque acting on it. 
Similarly, at point $L$, the source would be on the lower boundary at the $P \simeq 3$~s point if it were evolving without a disc.
The sources remaining above the upper border of the death valley when the disc becomes inactive (end points of the model curves) cannot reach even the $U$ point, so we do not see any of the four points along these curves (curve~5). For some sources, the lifetime with an active disc terminates at some point inside the valley. These sources cannot reach the lower border, and we see only the points $U$ and $U'$ on those model curves (curve~4). All four points ($U$, $U'$, $L$, and $L'$) appear only on the model curves with end points below the lower border (curves~1--3). There could be some exceptional cases for the sources that are born inside the valley.

\section{Results and Discussion} \label{results}

In the fallback disc model, RRATs are estimated to have evolutionary links with several other INS populations \citep{Gencali2024}. 
In Section~\ref{3.1}, we investigate the RRAT phases of different INS populations described in Section~\ref{the model} and compare our results with the observations. 
In Section~\ref{3.2}, we also analyze possible descendants of CCOs based on our results in the present and earlier work.

\subsection{RRAT phases of INSs} \label{3.1}

\begin{figure*}
    \centering
    \includegraphics[width=\linewidth]{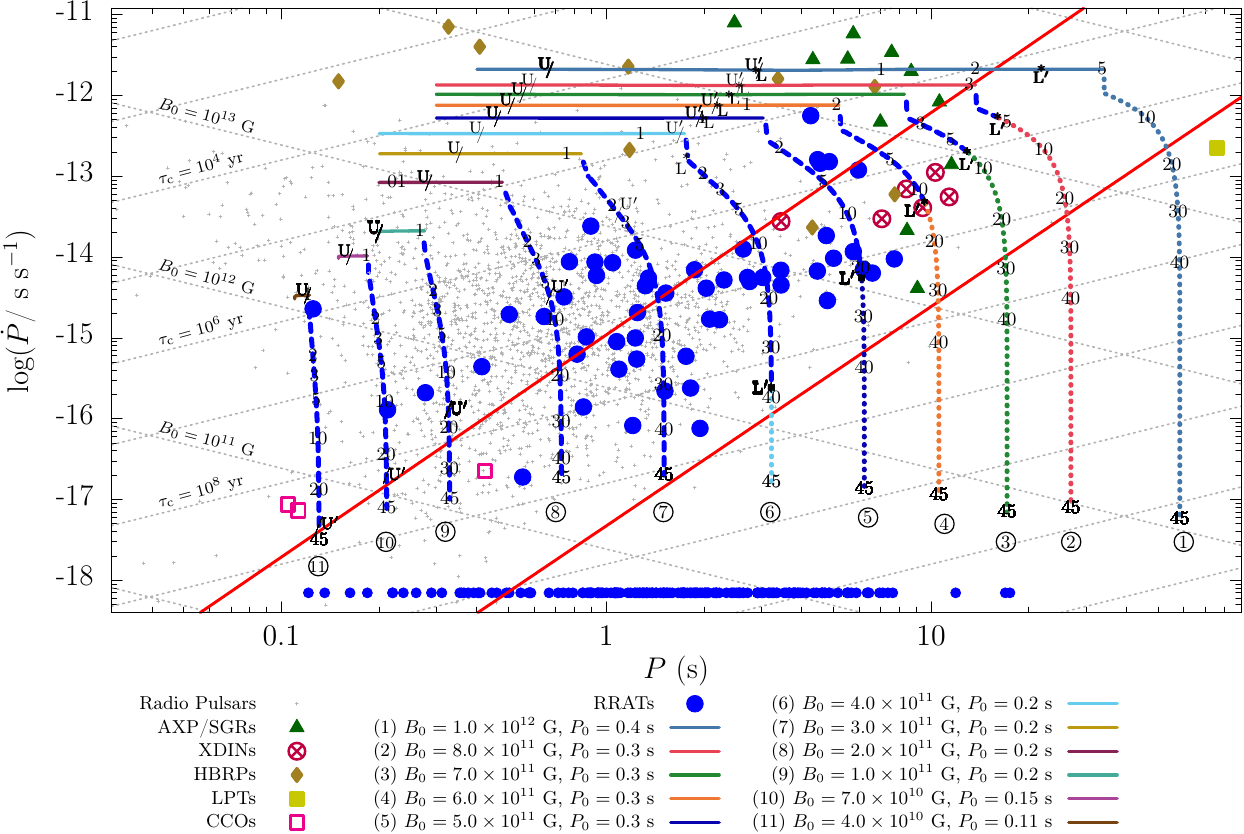}
    \caption{Illustrative model curves for the sources that start their evolution in the WP phase. 
    The lines and symbols are described in Fig.~\ref{fig:PPdot_model}.
    The model parameters are the same as given in Fig. \ref{fig:PPdot_model}. Initial conditions $B_0$ and $P_0$ are given below the figure.
    The data for RPs, CCOs, HBRPs, XDINs, AXP/SGRs, and LPTs are taken from the ATNF Pulsar Catalogue\protect\hyperlink{fn:ATNF}{\protect\footnotemark[2]}~\citep[version 2.7.0;][]{Manchester2005}, while the RRAT data are taken from \textsc{RRATalog}\protect\hyperlink{fn:rratalog}{\protect\footnotemark[3]}~\citep{Agarwal2026}.
    The small filled circles on the $P$ axis show $P$ values of RRATs with unknown $\Pdot$ values.}
    \label{fig:PPdot_WP}
\end{figure*}

\begin{figure*}
    \centering
    \includegraphics[width=\linewidth]{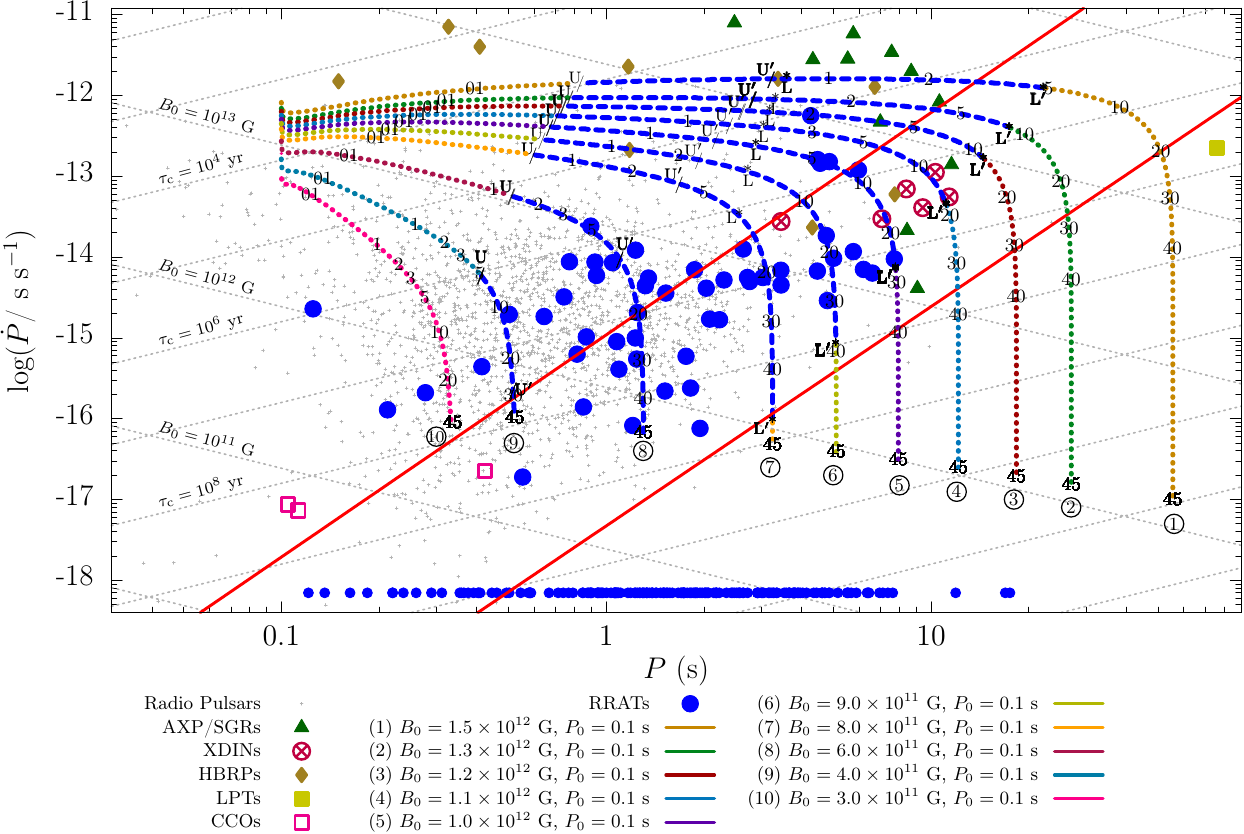}
    \caption{Illustrative model curves for the sources that start their evolution in the SP phase. 
    The lines and symbols are described in Fig.~\ref{fig:PPdot_model}.
    The model parameters are the same as given in Fig. \ref{fig:PPdot_model}. 
    Initial conditions $B_0$ and $P_0$ are given below the figure.
    See Fig.~\ref{fig:PPdot_WP} for references.
    }
    \label{fig:PPdot_SP}
\end{figure*}

Tracing the initial conditions ($\Md$, $P_0$, and $B_0$) estimated in our earlier work  \citep[see e.g.][]{Gencali2024}, we obtain the RRAT phases of INSs along the model curves seen in Figs \ref{fig:PPdot_WP}–\ref{fig:PPdot_CCO} (blue dashed segments).  
For all these model curves, we used disc parameters ($\alpha$, $\Tp$, and $C$) similar to those employed in our earlier work (given in the caption of Fig. \ref{fig:PPdot_model}).
\footnotetext[2]{%
\hypertarget{fn:ATNF}{}%
\url{https://www.atnf.csiro.au/research/pulsar/psrcat/}%
}
\footnotetext[3]{%
\hypertarget{fn:rratalog}{}%
\url{https://rratalog.github.io/rratalog}%
}

There are two main evolutionary paths depending on the initial conditions:
(1) The source enters a long lasting WP phase at an early stage of the evolution and transitions to the SP phase at a later time.
(2) The source always evolves in the SP phase without any accretion on to the NS.
The initial conditions leading to these different evolutionary avenues are discussed in detail in \citet{Gencali2024}. While the sources with relatively short $P_0$ tend to follow path (2),  the long-term rotational evolution depends most sensitively on $B_0$.
For a clear understanding of how the long-term evolution and the accompanying RRAT phase change with $B_0$, we present the results of our simulations separately for cases (1) and (2) in Figs~\ref{fig:PPdot_WP} and \ref{fig:PPdot_SP} respectively.

In Fig.~\ref{fig:PPdot_WP}, all model sources start their evolution in the WP phase (solid lines). For these model curves, except curve 1, the RRAT phase starts immediately after the WP/SP transition (at the end of the solid curves). For curve 1, there is no RRAT phase, because the source crossed the $L'$ point during the WP phase. In this phase, continuous accretion on to the NS does not allow RRAT or ordinary pulsed radio emission. It is seen that the evolutionary curves with higher $B_0$ reach longer $P$ and have shorter RRAT phase. For sources with $B_0 \gtrsim 10^{12}$ G, there is no RRAT phase. With few exceptions, all known RRATs have $P < 10$ s, which is the maximum $P$ reached by the model source 4 with $B_0 \simeq 6 \times 10^{11}$ G.
The region formed by the RRAT phases (dashed curves) extends from CCOs to XDINs, leaving persistent AXP/SGRs and LPTs, which have relatively high $B_0$, outside the region  \citep[see][for a detailed discussion on possible evolutionary links between the INS systems]{Gencali2024}.
It is seen in Fig.~\ref{fig:PPdot_WP} that the RRAT region estimated by our calculations covers most of the observed RRATs, with few exceptions, and seems to be consistent with the borders of the observed RRAT distribution.
Along the evolutionary paths 1--5, the sources never enter a RP phase since their $L$ points are reached while the sources are still evolving in the WP phase. For the curves 7--11, the sources could be RPs after their WP/SP transitions, and could become RRATs at any point along their RRAT phases (dashed curve). It is also possible that some of these sources remain as RPs even after the inactivation of the disc depending on the actual location of their death point inside the death valley.

The model curves in Fig.~\ref{fig:PPdot_SP} illustrate the evolution of the sources that evolve always in the SP phase. Compared to Fig.~\ref{fig:PPdot_WP}, 
the RRAT phases last considerably longer and cover a wider range of $P$ along each track. Note that the time intervals between $U$  and $U'$ for the curves 1--7 are rather short ($\approx 10^5$~yr) which indicates a low probability of observing RRATs in this region. Indeed, the ages of most RRATs  estimated in our model are $\gtrsim 5 \times 10^5$~yr. 

Details of the initial conditions reproducing the individual source properties ($\Lx$, $P$, and $\Pdot$) for all INS populations and possible evolutionary connections between them were studied in our earlier work (see Sec. \ref{intro}). 
Here, we would like to stress the evolutionary link between RRATs and XDINs. Our results in the present and earlier work indicate that RRATs with relatively strong dipole fields evolve into the XDIN region.
In this model, XDINs are estimated to be evolving in the SP phase with $B_0$ values placing them below the lower border of the death valley \citep{Ertan2014}. 
Their discs are still active while the disc enhancement is not sufficient to yield RRAT bursts. 
From the evolutionary curves passing through the XDIN region (Figs \ref{fig:PPdot_WP} and \ref{fig:PPdot_SP}), we estimate that XDINs evolved in the RRAT phases  before they appear with the XDIN properties. 
It is seen in Fig.~\ref{fig:PPdot_WP} that the model curves leading to the longer $P$ values have shorter RRAT epochs, constraining the RRAT behavior to smaller fractions of the sources with death points closer to the lower border of the valley. In other words, the probability of having an RRAT episode decreases for the sources evolving towards the XDIN region (to longer periods).
This could be the physical reason for the observed cut-off of the P distribution of RRATs. 
In the model, the evolution of the sources with low $P$ values ($P \lesssim 0.6$~s) is sensitive to their $P_0$, as well as $B_0$, and should be investigated separately (see Sec. \ref{3.2}). 
RRATs with the lowest $P$ values are more likely to start their evolution in the WP phase (see Fig.~\ref{fig:PPdot_CCO}).

\subsection{Descendants of CCOs} \label{3.2}

Descendants of CCOs are the subject of a long-lasting problem in the frame of the models assuming that CCOs evolve purely by magnetic dipole torques in vacuum \citep{Vigano2013hiddenB,Gotthelf2013DRP,Gotthelf2013CCOs,Luo2015}.
The dipole moments, rotational and $\Lx$ evolution of these sources are very different in the dipole torque and the fallback disc models. Below, for comparison, we first summarize the predictions of the dipole torque model and the results of the related observations. Next, we  describe the evolutionary links and possible descendants of CCOs estimated in the fallback disc model based on our results obtained in Sec. \ref{3.1} for the RRAT phase. 

\begin{figure*}
    \centering
    \includegraphics[width=\linewidth]{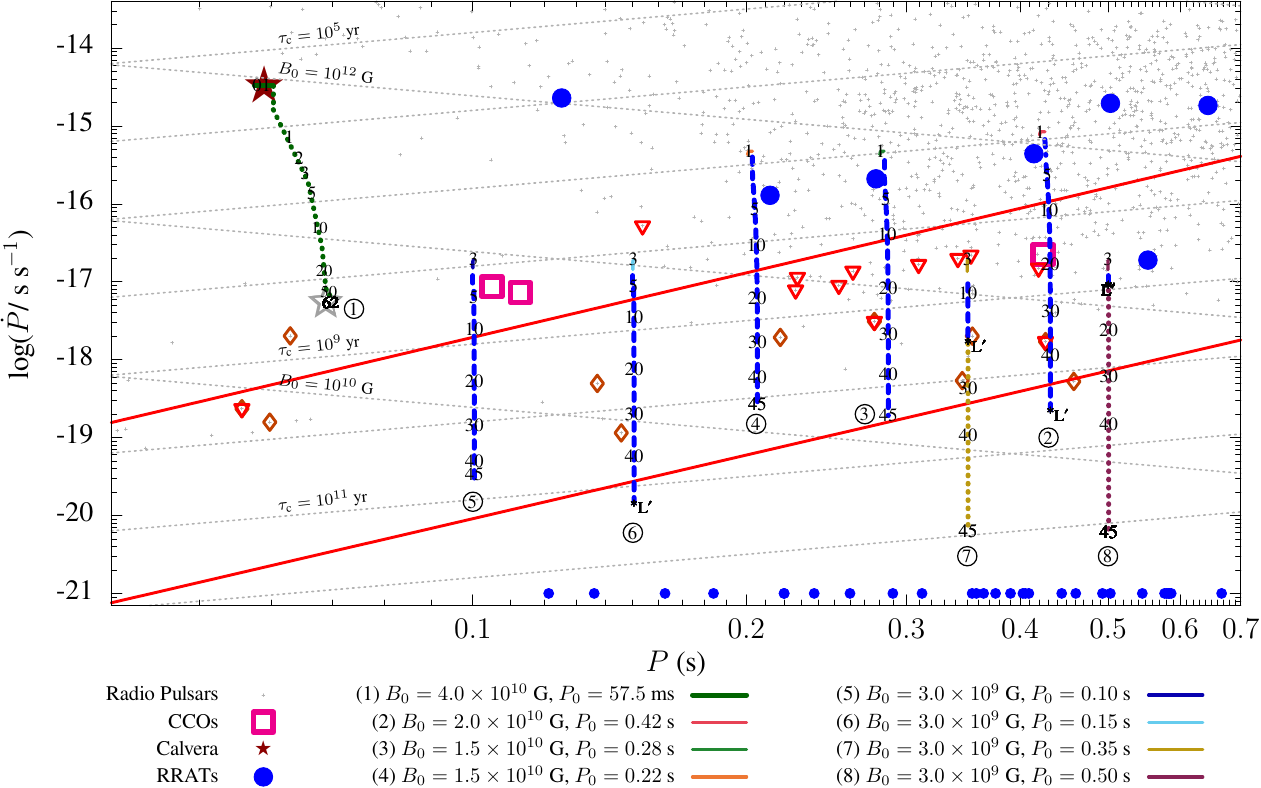}
    \caption{Illustrative model curves for CCO evolution. 
    The lines and symbols are described in Fig.~\ref{fig:PPdot_model}.
    The model parameters are the same as given in Fig. \ref{fig:PPdot_model} with difference in $\eta=0.6$ (curve 1), $\Md \simeq 5 \times 10^{-7} \Msun$ (curve 1) and $\Md \simeq 6.3 \times 10^{-7} \Msun$ (curves 5--8). 
    Initial conditions $B_0$ and $P_0$ are given below the figure. 
    The empty diamonds and triangles indicate the RPs investigated in X-rays as candidate immediate descendants of CCOs by \citet{Gotthelf2013DRP} and \citet{Luo2015}, respectively.
    The filled and empty stars on curve 1 show the current position and the final location of "Calvera" at the end of its evolution in the model, respectively.
    See Fig.~\ref{fig:PPdot_WP} for references.
    }
    \label{fig:PPdot_CCO}
\end{figure*}

The properties of CCOs estimated in the dipole torque model can be summarized as follows. 
The $B_0$ values inferred from the dipole torque formula are in the $(6-20)\times 10^{10}$~G range 
for the three CCOs with measured $P$ and $\Pdot$ values. For these sources, $P$ values ($\sim 100 - 400$~ms) are sufficiently short to produce pulsed radio emission.  
Nevertheless, no regular radio pulsations were observed from about 10 confirmed CCOs \citep{DeLuca2017}.
Recently, radio pulses were detected from a CCO, 1E 1207.4-5209, while the pulsed radio emission of the source seems to be transient, not regular \citep{Zhang2026}. 
It was proposed that the lack of radio pulsations could be due to the beaming effect \citep{Halpern2010}. 
In this case, some immediate descendants of CCOs could be among the RPs that could still be detected in X-rays produced by their cooling. 
With the dipole-torque assumption, it is also possible that the pulsed radio luminosities of these sources with their typical rotational power $\Edot \sim 10^{31}$--$10^{32}$~\ergpers~could remain below the detection limits of radio observations \citep{Faucher2006, Johnston2020, Sautron2024, Pardo2025}. 
Assuming that some of these sources could have  weaker dipole fields ($B_0 < 6 \times 10^{10}$ G), $14$ RPs (empty diamonds in Fig. \ref{fig:PPdot_CCO}) that satisfy this condition were observed but not detected in the X-ray band \citep{Gotthelf2013DRP}.
The ages of these sources inferred from the $\Lx$ upper limits are $>10^4$–$10^5$~yr, which are substantially greater than those of CCOs under standard cooling scenarios. 
These results indicate that these sources are unlikely to be immediate descendants of CCOs \citep{Gotthelf2013DRP}.

An alternative idea about CCOs is that their dipole fields could be initially buried into the crust by the fallback SN matter, and could be growing during the subsequent evolution  \citep{Geppert1999, Geppert2013, Ho2011, Vigano2013hiddenB}. That is, immediate descendants of some of CCOs could be RPs with $B_0$ values 
greater than those of newly born CCOs 
\citep{Gotthelf2013DRP, Luo2015}. 
To test this prediction, 12 RPs with $B_0 \sim 10^{11}$~G were observed and could not be detected in  X-rays \citep{Luo2015}. 
This means that these 12 RPs (empty triangles in Fig. \ref{fig:PPdot_CCO}) cannot be the immediate descendants of CCOs either. 
In this dipole torque model, statistical studies indicate that beaming alone is unlikely to account for the radio non-detection of all CCOs \citep{lu2024upperlimitsradiopulses}. 
Then, the absence of estimated immediate descendants of CCOs in radio surveys might indicate that CCOs could be intrinsically radio-quiet due to an unknown physical mechanism in this model \citep{Gotthelf2013DRP, Luo2015,DeLuca2017}.

In the fallback disc model, the picture is completely different. Based on the results from our earlier and the present work, we could summarize the properties of CCOs and their possible descendants as follows.
The properties of CCOs can be reproduced with relatively low  $B_0$ ($\sim (2$–$5) \times 10^9$~G), when the sources are evolving in the WP phase \citep{Benli2018CCOs}. 
In this phase, mass accretion on to the magnetic poles of the star suppresses radio emission, which is the physical reason for the absence of radio pulsations from CCOs in this model. 
These sources are expected to make a transition from the WP to the SP phase at ages of $\sim 10^5$~yr. After this transition, they  could show pulsed radio emission if they have currently sufficient rotational power.

In Fig.~\ref{fig:PPdot_CCO}, the model curves (5--8) show the examples of the evolutionary paths that could represent the evolution of the sources with initial rotational properties similar to the current properties of the three CCOs with $0.1 < P < 0.5$~s and $\Pdot \sim 10^{-17}$~\spers.  
The $P$ values of all these model sources remain close to their $P_0$ values during their long-term evolution. 
These sources are not likely to emerge as RPs after the WP/SP transition (close to the initial points of the evolutionary curves in the \PPdot~diagram). 
Only the sources with the shortest $P$ values could become RPs provided that their death points are close to the lower border of the death valley. 
For instance, the source represented by curve 5 can become a RP after the WP/SP transition if its death point is below the end point of its evolutionary curve. 
For the model curves 2--6, the RRAT phase terminates with the inactivation of their discs. 
For the sources 7 and 8, the RRAT phase ends while their disc is still active. These sources are not likely to be detected as X-ray or radio sources after their RRAT bursts are switched off.

The recently identified CCO Calvera (filled star in Fig.~\ref{fig:PPdot_CCO}) has a relatively short $P$ ($\simeq 59.2$~ms) and a high $\Pdot$ ($\simeq 3.2 \times 10^{-15}$~\spers) compared with the three CCOs with measured $\Pdot$ \citep{Rigoselli2024}. For the illustrative model curve representing the evolution of Calvera (curve~1 in Fig.~\ref{fig:PPdot_CCO}), we take $B_0 = 4.0 \times 10^{10}$~G, $P_0 = 57.5$~ms, $M_{\rm d} \simeq 5 \times 10^{-7}$~M$_{\odot}$, and $\eta = 0.6$ \citep{Gencali2026}. Illustrative evolutionary curves for a large range of initial conditions ($3 \times 10^{9} < B_0 < 4 \times 10^{10}$~G and $25 < P_0 < 55$~ms) similar to those of Calvera are given in \citet{Gencali2026}.
Similar to other CCOs, its lack of radio pulsation is due to the accretion on to the NS in the WP phase in this model.
Unlike the three CCOs, the disc of Calvera becomes inactive above the death valley (empty star in Fig. \ref{fig:PPdot_CCO}). 
This model source becomes an RP after the WP/SP transition and remains so for a long time ($ > 10^8$~yr). There are $\sim 15$ RPs in the region where old descendants of Calvera-like sources are located. 
In the model, this low number of old descendants indicates that the birth rate of Calvera-like sources is very low. 
This is supported by the lack of no CCOs detected in the region around Calvera in the \PPdot~diagram. 
From these model results, we estimate that immediate descendants of CCOs are not likely to be observed as RPs (see \citet{Gencali2026} for a detailed investigation of the evolution of the Calvera like sources).

\section{Conclusions} \label{conclusions}

We have investigated the idea proposed by \citet{Gencali2024} that the RRAT mechanism may arise from the enhancement of the voltage across the polar caps of dead pulsars caused by the penetration of the fallback disc into the light cylinder in the SP phase. 
We estimate that the transient nature and short duration of the radio bursts could be due to continual opening and reconnection cycle of the field lines at the inner disc boundary on the dynamical timescale,  
which could intermittently yield the conditions required for radio emission. 
We have found that the RRAT region and its borders predicted in the model seem to be in good agreement with the distribution of observed RRATs. Our model does not address the details of the RRAT emission mechanism. Nevertheless, our results indicate that the disc--magnetosphere interaction could be responsible for the RRAT emission.
From our earlier simulations, we estimate that most RRATs have low X-ray luminosities powered by the cooling of the NS \citep{Gencali2021RRATs},
which is  consistent with non-detection of most RRATs in X-rays.
Our results  support the earlier prediction by \citet{Gencali2024} that XDINs passed through RRAT phases in earlier phases of their long-term evolution with duration $\lesssim 10^6$~yr.

We have also investigated the evolution and descendants of CCOs in the same model. 
In our model, the CCOs similar to the three CCOs with measured $P$ and $\Pdot$, are not likely to become RPs, while they could become RRATs with lifetimes $\sim 10^6$~yr.
The situation is different for CCOs with lower $P$ and/or higher $\Pdot$ values than those of the three CCOs, like Calvera.
We estimate that the discs of Calvera-like sources become inactive while the sources are evolving above the death valley. 
As a result, they are likely to become RPs after accretion on to the NS terminates at ages $t \gtrsim 10^{4}$~yr. 
They remain as RPs for long timescales ($\gtrsim 10^{8}$~yr). 
There is indeed a cluster of RPs in the region hosting the descendants of Calvera-like CCOs in our model (see \citet{Gencali2026} for details). 
In this picture, the lack of Calvera-like CCOs and the small number of observed RPs that could be old descendants of these sources imply that these sources have a low birth rate, which also suggests that their immediate descendants may not exist in the Galaxy at present. 

The presence of fallback discs around four CCOs was tested by the optical and IR observations, and no detections were reported \citep{Wang2007remnants}. Note that the optical/IR emission of the disc is powered by the X-ray irradiation. These non-detections are consistent with the relatively low X-ray luminosities of these sources, and do not exclude the presence of fallback discs. Detailed modelling showed that the optical and mid-IR emission of 4U 0142+61 can be reproduced by an irradiated, active fallback disc \citep{Ertan2006, Ertan2007}. Deep IR observations, particularly with JWST, of the nearby CCOs and RRATs at their young ages would provide critical tests for the fallback disc model, and could constrain their disc properties.

\section*{Acknowledgements}

We thank the referee, Matteo Sautron, for the meticulous review and very useful comments on the manuscript that significantly improved our work.
We acknowledge research support from Sabanc{\i} University, and from T\"{U}B\.{I}TAK (The Scientific and Technological Research Council of Turkey) through grant 125F197. 

\section*{Data Availability}
No new data were analysed in support of this paper.




\bibliographystyle{mnras}
\bibliography{mnras} 







\bsp	
\label{lastpage}
\end{document}